\documentclass[twocolumn,secnumarabic,amssymb, nobibnotes, aps, prd]{revtex4}

\usepackage{graphicx}

\usepackage{epstopdf}

\begin{document}

\title{Oxygen doping of P3HT:PCBM blends: Influence on trap states, charge carrier mobility and solar cell performance}

\author{Julia Schafferhans $^1$}\email[Electronic mail: ]{julia.schafferhans@physik.uni-wuerzburg.de}
\author{Andreas Baumann $^1$}
\author{Alexander Wagenpfahl $^1$}
\author{Carsten Deibel $^1$}
\author{Vladimir Dyakonov $^{1,2}$}
\email[Electronic mail: ]{dyakonov@physik.uni-wuerzburg.de}
\affiliation{$^1$ Experimental Physics VI, Faculty of Physics and Astronomy, Julius-Maximilians-University of W\"urzburg, Am Hubland, 97074 W\"urzburg, Germany}
\affiliation{$^2$ Bavarian Center of Applied Energy Research e.V. (ZAE Bayern), Am Hubland, 97074 W\"urzburg, Germany}

\begin{abstract}
We investigated the influence of oxygen on the performance of P3HT:PCBM (poly(3-hexylthiophene):[6,6]-phenyl C61 butyric acid methyl ester) solar cells by current--voltage, thermally stimulated current (TSC) and charge extraction by linearly increasing voltage (CELIV) measurement techniques. The exposure to oxygen leads to an enhanced charge carrier concentration and a decreased charge carrier mobility. Further, an enhanced formation of deeper traps was observed, although the overall density of traps was found to be unaffected upon oxygen exposure. With the aid of  macroscopic simulations, based on solving the differential equation system of Poisson, continuity and drift-diffusion equations in one dimension, we demonstrate the influence of a reduced charge carrier mobility and an increased charge carrier density on the main solar cell parameters, consistent with experimental findings.

\end{abstract}

\maketitle

\section{Introduction}
\label{Introduction}

Power conversion efficiencies of almost 8\% for organic solar cells have already been achieved, with growing tendency
\cite{greenV35}. A critical issue yet to be addressed are the factors influencing the device lifetime. To gain a detailed understanding 
of the device stability, as well as the underlying degradation mechanisms and their impact on the solar cell performance is an important prerequisite for lifetime enhancements. 

Organic solar cells undergo many degradation pathways during their lifetime. Efficiency losses due to light \cite{reese2008}, oxygen 
\cite{seemann2009, neugebauer2000} and water \cite{norrman2009} are reported. The details of the degradation processes, 
however, are still not completely understood. Other studies, on the contrary, are only focused on the degradation of a single active 
material instead of the blend. For example for P3HT oxygen is known to form a charge transfer complex \cite{abdou1997} resulting 
in p-doping of P3HT \cite{liao2008, meijer2003}, which is also investigated theoretically by band-structure calculations \cite{lu2007}. 
Furthermore, oxygen induced degradation of P3HT is reported to result in decreased mobilities \cite{abdou1997, schafferhans2008} 
and increased trap densities \cite{schafferhans2008}. In the case of C60, oxygen also causes decreased electron mobilities 
\cite{tapponnier2005, koenenkamp1999} and increased trap densities \cite{matsushima2007}, as demonstrated by investigations of C60 
based field effect transistors.

This paper addresses the oxygen induced degradation of poly(3-hexylthiophene):[6,6]-phenyl C61 butyric acid methyl ester solar 
cells in the dark, as well as under simultaneous illumination. Detailed investigations including charge carrier mobility, charge carrier 
density and defect spectroscopic measurements are performed to get an insight of the degradation mechanism due to oxygen in 
P3HT:PCBM solar cells.

\section{Experimental}
\label{Experimental}
The investigated bulk heterojunction solar cells were prepared by spin coating P3HT:PCBM (ratio 1:0.8) blends made from solutions 
of 30 mg/ml in chlorobenzene on  poly(3,4-ethylenedioxythiophene) poly(styrenesulfonate) (Pedot:PSS) covered indium tin oxide/glass 
substrates. The active layer was about 240 nm thick. After an annealing step of 10 min at 130 $^\circ$C, Ca (1.5nm)/Al (60nm) 
contacts were evaporated thermally (base pressure during evaporation $< 7\cdot 10^{-7}$  mbar). The effective areas of the solar cells were 3 mm$^2$ and 9 mm$^2$.

PEDOT:PSS was purchased by H.C.Starck (CLEVIOS P VP Al 4083). P3HT was delivered by BASF (Sepiolid P200), PCBM (purity 99,5 \%) by Solenne. All materials were 
used without further purification. All preparation steps were performed in a nitrogen glovebox and an attached thermal vacuum 
evaporation chamber. 

Initial current--voltage (IV) characteristics of the solar cells were measured in the nitrogen glovebox. For the illumination of the cells 
we used an Oriel 81160 AM1.5G solar simulator. The power conversion efficiencies of the investigated samples were  3\% -- 3.5\%.

Degradation studies of the unprotected solar cells were performed in a closed cycle cryostat. During the thermally stimulated current (TSC), charge carrier 
extraction by linearly increasing voltage (CELIV) \cite{juska2000, deibel2009a} and IV--measurements the cryostat was filled with helium as 
contact gas.  To investigate the influence of oxygen on the solar cells, the helium was replaced with synthetic air (80\% N$_2$, 20\% 
O$_2$, $<$ 1ppm H$_2$O) for different degradation times. Thus, the effect of humidity was excluded. This exposure was always done at a constant temperature of 300 K. Two 
different degradation conditions were investigated: in synthetic air either in the dark or under illumination. For  the latter we used a 
10 W high power white light emitting diode (Seoul). The light intensity was adjusted to match the short circuit 
current of the non degraded cells obtained under the solar simulator. The LED was also used for the illuminated IV-curves in the cryostat. 

For TSC measurements the trap filling was achieved by illumination of the samples at 18 K for five minutes using the 10 W LED. 
After a dwell time of five minutes the temperature was increased with a constant heating rate of 7 K/min up to 300 K. The TSC 
signals were detected by a Sub-Femtoamp Remote Source Meter (Keithley 6430) without applying an external field, implying that 
the detrapped charge carriers were extracted from the samples only due to the built-in voltage.  Further details of the TSC 
measurements are described elsewhere \cite{schafferhans2008}. 

CELIV measurements were performed at 300 K. The peak voltage of the triangular bias pulse in reverse direction was V$_a$ = 2 V 
applied on the Ca/Al electrode.  A voltage pulse width of 50 $\mu$s was chosen. 

\section{Results and Discussion}
\label{Results and Discussion}

\subsection{Experimental Results}
\label{Experimental Results}
The IV-curves for different degradation times
\begin{figure}[htb]
	\centering
	\includegraphics[width=6cm]{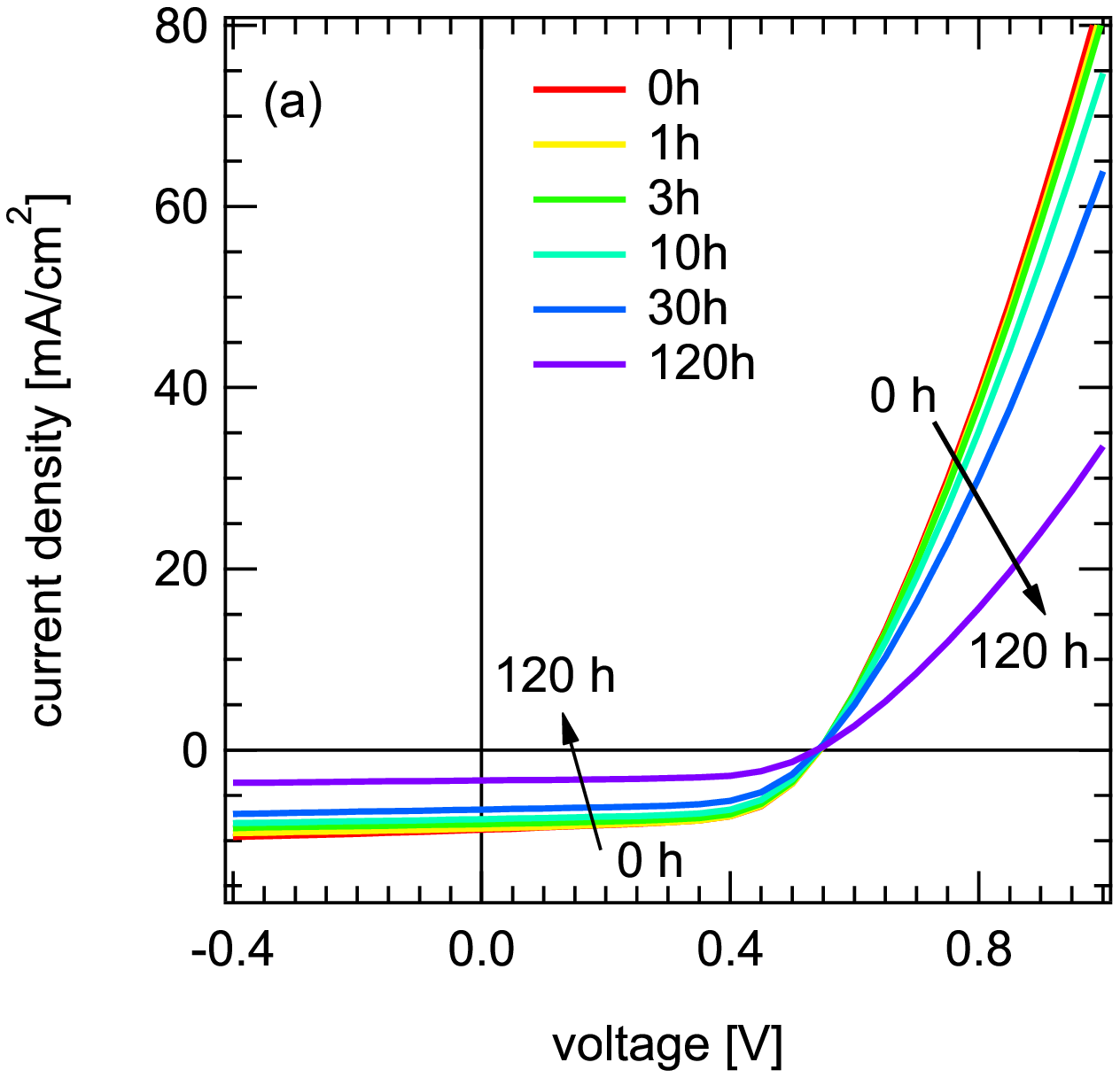}
	\includegraphics[width=6cm]{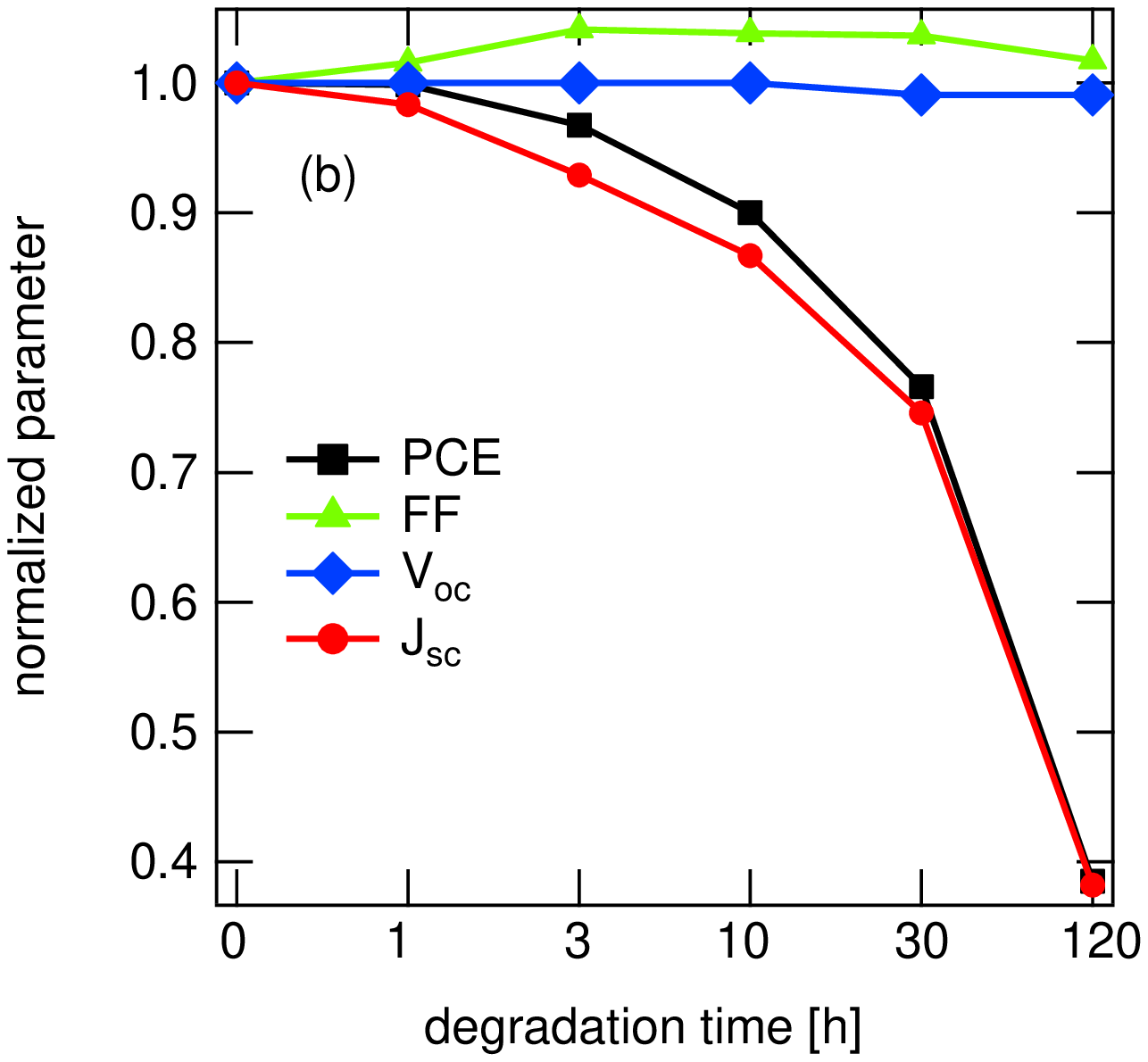}	
	\caption{Illuminated IV-curves of P3HT:PCBM solar cells for different dark degradation times in synthetic air (a) and the 
normalized values of the PCE, fill factor, open circuit voltage and short circuit current (b). The efficiency loss is due to the decrease of 
the short circuit current.}
	\label{fig:Fig1}
\end{figure}
 between 0 h and 120 h in synthetic air in the dark are shown in Fig. \ref{fig:Fig1}a. The 
obtained values for the power conversion efficiency (PCE), fill factor (FF), open circuit voltage (V$_{oc}$) and short circuit current 
(J$_{sc}$), normalized to the initial values are presented in Fig. \ref{fig:Fig1}b
The observed efficiency loss with dark degradation time is only due to J$_{sc}$, which decreases about 60\% within 120 h, whereas 
V$_{oc}$ and FF remain almost constant.
Since the exposure of the samples to synthetic air in the dark results only in a loss in J$_{sc}$, a degradation of the electrodes being the origin of these findings can be excluded. Degradation (oxidation) of the Ca/Al electrode would result in the occurrence of an s-shaped IV-curve \cite{wagenpfahl2010}, at least in a strong loss in FF or V$_{oc}$. Neither of these effects was observed within the timescale of degradation considered here. 
\begin{figure}	[htb]
	\centering
	\includegraphics[width=6cm]{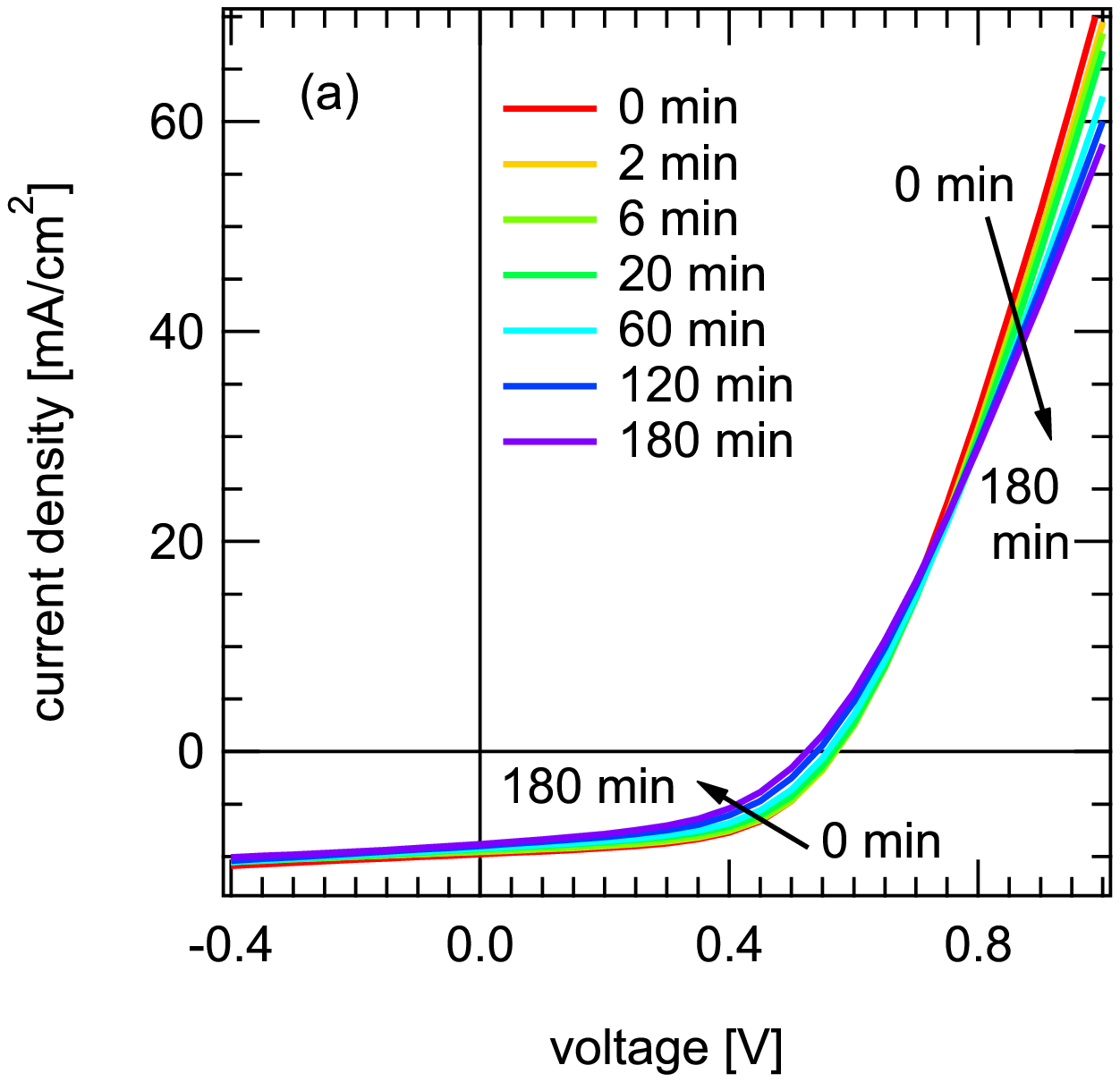}
	\includegraphics[width=6cm]{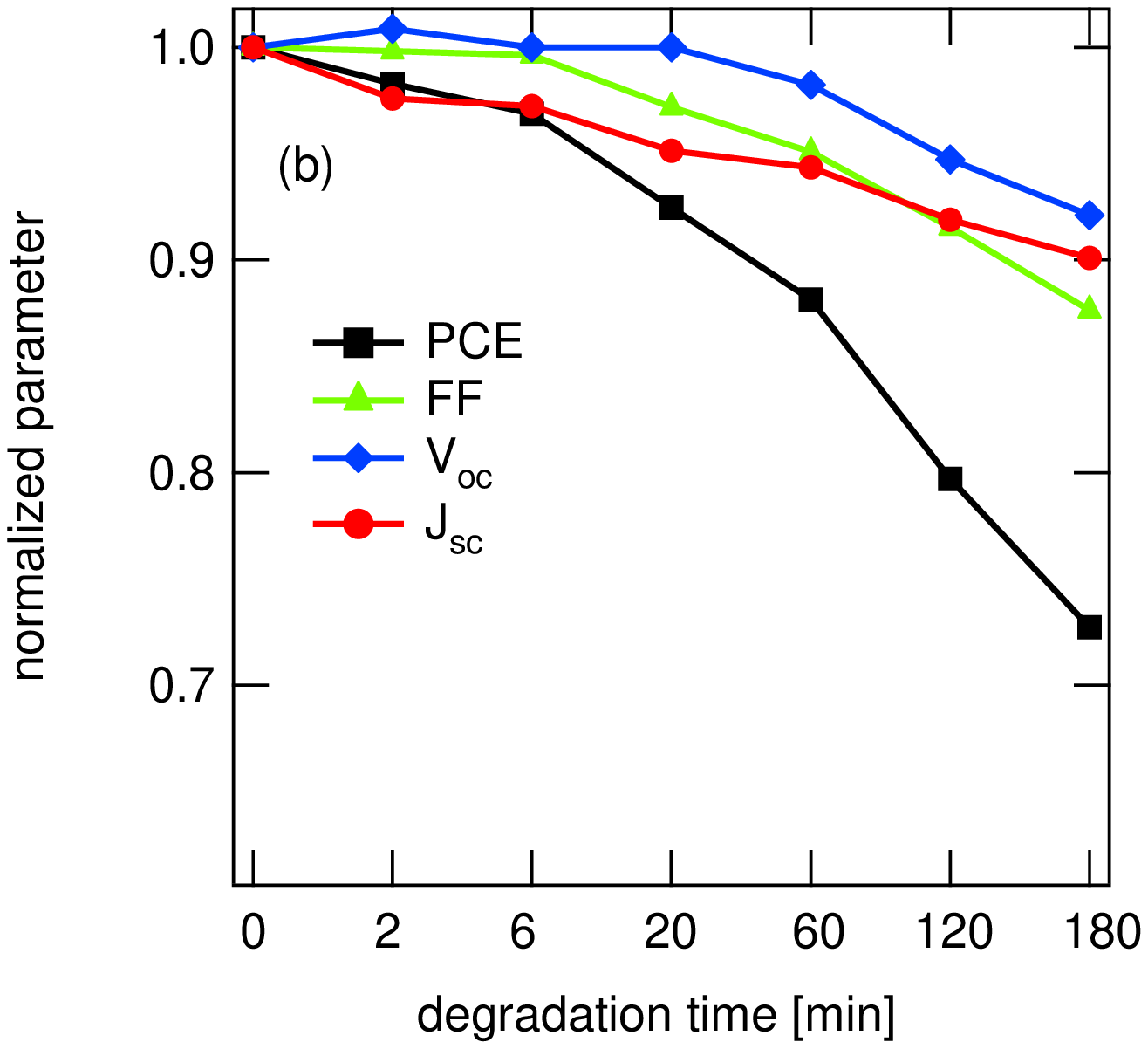}	
	\caption{Illuminated IV-curves of P3HT:PCBM solar cells for different degradation times in synthetic air under simultaneous illumination (a) and the normalized values of the PCE, fill factor, open circuit voltage and short circuit current (b). In contrast to 
dark degradation in synthetic air all parameters decrease.}
	\label{fig:Fig2}
\end{figure}

In contrast to degradation in synthetic air in the dark, the efficiency loss due to degradation in synthetic air and simultaneous 
illumination (Fig. \ref{fig:Fig2}a, b) occurs on a faster time scale---about 30\% efficiency loss within 180 min---and is caused by a decrease of J$_{sc}
$, as well as FF and V$_{oc}$, which all decrease about 10\% within this time scale. Since, in the case of dark degradation, FF and V$_{oc}$ are unaffected, the decrease of both during photodegradation is due to light in the presence of oxygen, indicating an additional degradation path due to photooxidation in addition to doping as found under oxygenation in the dark.

\begin{figure}[htb]
	\centering
	\includegraphics[width=6.5cm]{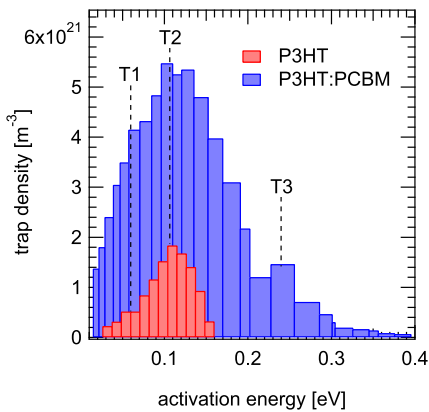}		
	\caption{Comparison of the DOOS distribution of P3HT and P3HT:PCBM blends, as obtained by TSC-T$_{start}$-T$_{stop}$ 
measurements. The DOOS of P3HT consists of two different traps (T1, T2), which both also contribute to the DOOS of the blend. Additionally to these two traps the DOOS of the blend features a third trap distribution (T3) with higher activation energy.}
	\label{fig:Fig3}
\end{figure}

To gain a deeper understanding of the underlying degradation mechanisms, we performed TSC measurements to obtain information 
about the electronic trap states of P3HT:PCBM solar cells for dark as well as photodegradation.
First of all, the trap distribution of non degraded P3HT:PCBM solar cells was investigated by applying a fractional TSC 
measurement---the so-called T$_{start}$-T$_{stop}$ technique \cite{schafferhans2008, schmechel2004}---with T$_{stop}$ varied 
between 25 K and 150 K in steps of 5 K. The resulting density of occupied states (DOOS) distribution is shown in Fig. \ref{fig:Fig3}.  
The continuous activation energies of the traps range from 20 meV to 400 meV with the center of distribution at about 105 
meV. For further interpretation of the DOOS distribution of the blend, we consider the results for pure P3HT. For P3HT, the DOOS was related to two different overlapping traps with approximately Gaussian energy distributions 
\cite{schafferhans2008}, with the center of distribution of the dominant trap at about 105 meV (T2) and the other at about 50 meV (T1) (shoulder at the low energy side of the DOOS distribution) (Fig. \ref{fig:Fig3}). Since the centers of distribution for the blend as well as the pure P3HT are at 105 meV, we attribute the dominant trap to P3HT, although the distribution of the main trap in the blend is  broadened as compared to the pure P3HT, indicating a higher disorder in the blend. Furthermore, the T1 can also be seen in the DOOS distribution of the blend in the steep onset of the distribution on the low energy side, but is less pronounced due to the higher concentration of the main trap T2, overlapping T1. Additionally to these two traps, the DOOS distribution of the blend features further trap states  with activation energies of 200 meV to 400 meV which are not seen in pure P3HT, with the center at about 250 meV. We refer to the latter trap distribution as T3. Therefore we conclude that the DOOS of the blend consists of three different traps, with the dominant trap T2 at 105 meV. This interpretation is also supported by the TSC spectrum of   
pristine P3HT:PCBM solar cells, where T2 has a lower concentration compared to T1 and T3, so that T1, T2, and T3 can be clearly distinguished (data not shown).   
The trap density obtained from the TSC measurements of the investigated P3HT:PCBM solar cells is in the range of 6 -- 8 $\cdot 
10^{22}$ m$^{-3}$ which is considerably higher than the value obtained for P3HT (1 $\cdot 10^{22}$ m$^{-3}$) \cite{schafferhans2008}.

\begin{figure}[htb]
	\centering
	\includegraphics[width=6.0cm]{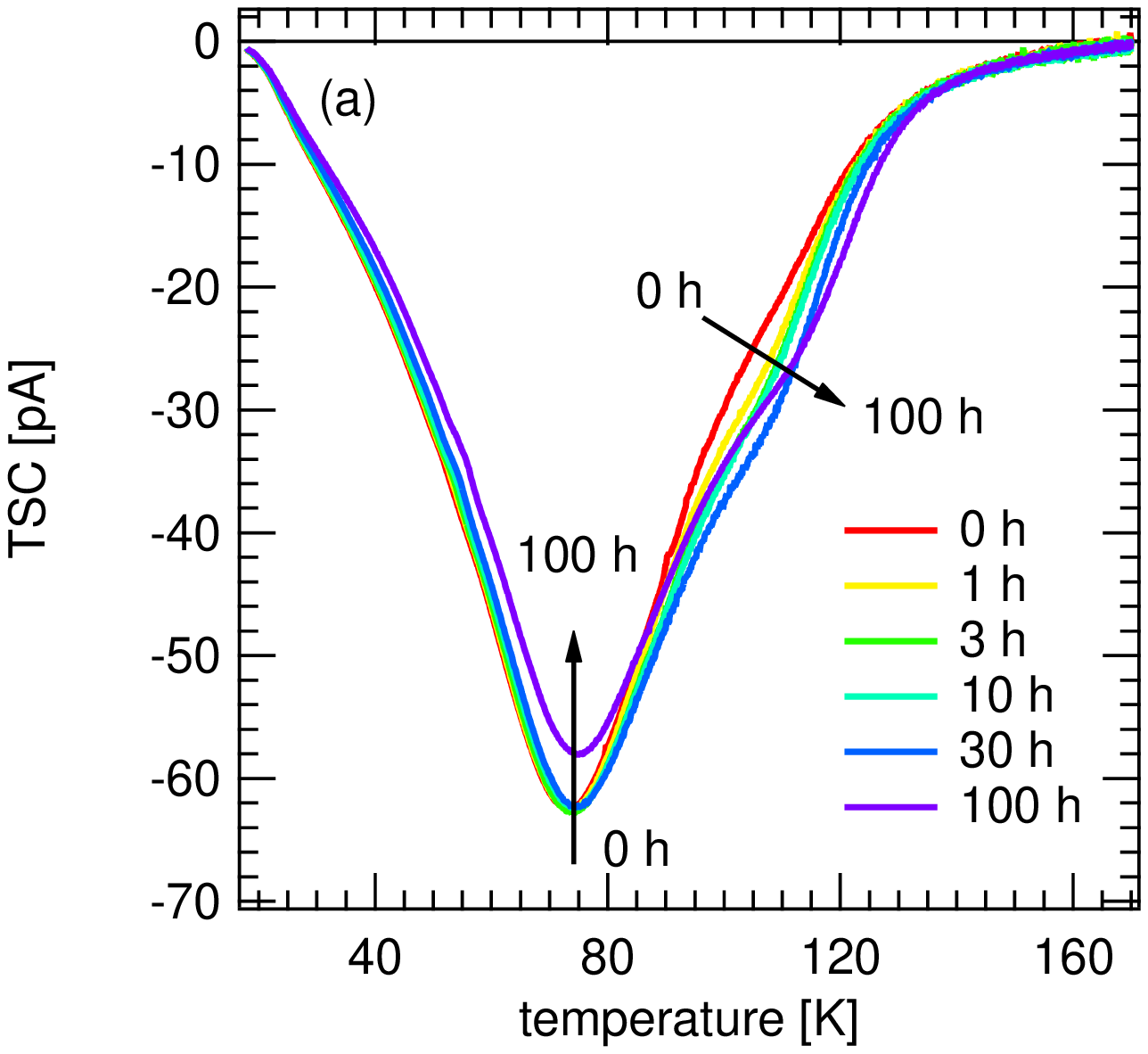}	
	\includegraphics[width=6.3cm]{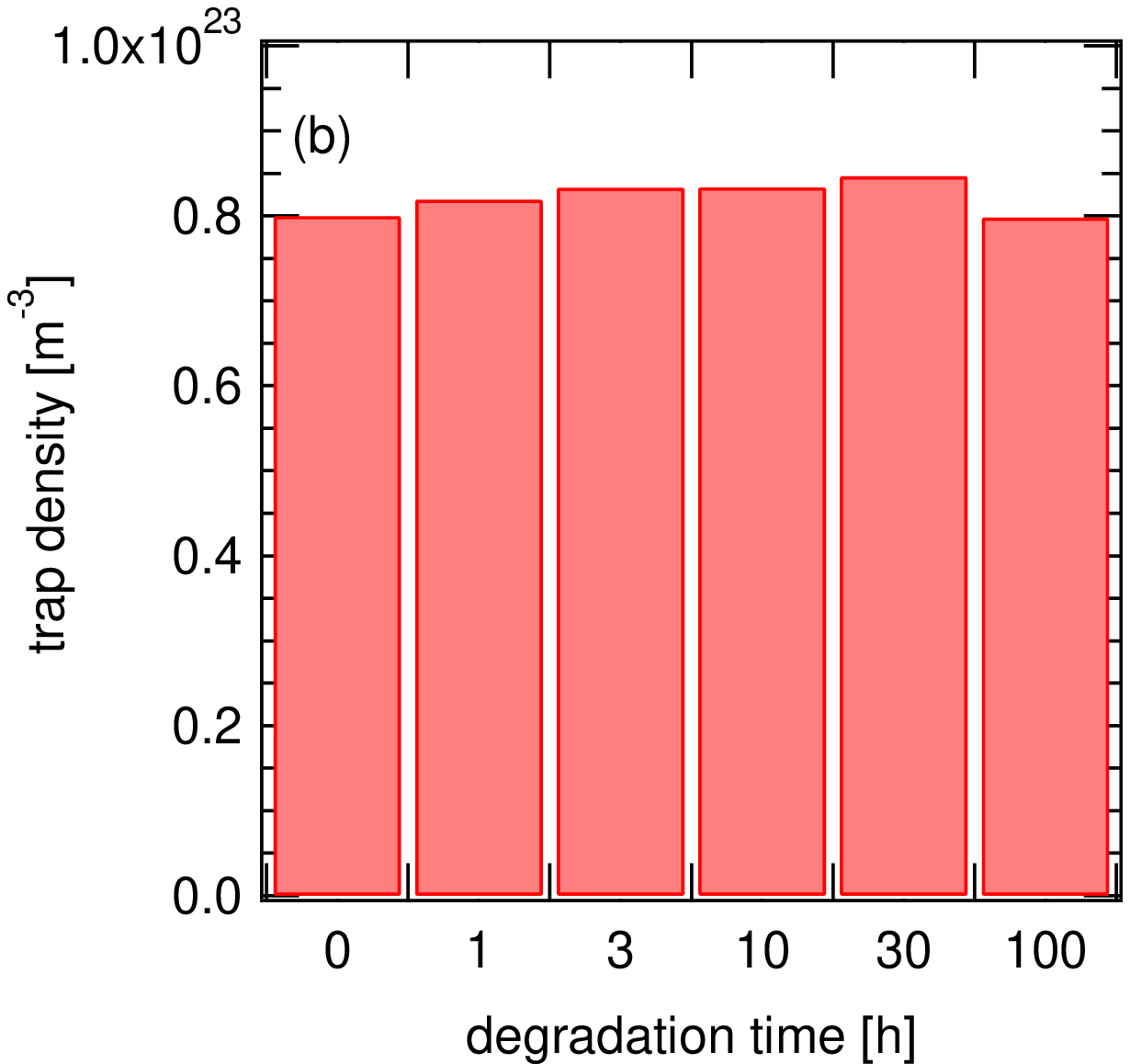}	
	\caption{TSC spectra of P3HT:PCBM solar cells for different dark degradation times in synthetic air (a) and the obtained trap 
densities (b). The trap density remains almost constant. The slight decrease of the trap density for 100 h degradation time is possibly 
due to recombination of the detrapped charge carriers.}
	\label{fig:Fig4}
\end{figure}

The TSC spectra of P3HT:PCBM solar cells for different dark degradation times in synthetic air are shown in Fig. \ref{fig:Fig4}a. 
Thermally stimulated current is observed between 18 K and 160 K. At higher temperatures no additional TSC peaks can be observed up to 300 K. With exposure to synthetic air, a shoulder at about 100 K appears, which increases with longer degradation times within the investigated time scale up to 100 h. The appearance of a shoulder indicates an increase in density of the deeper traps or even the formation of additional traps. The main peak however decreases slightly.
As a result the overall trap density of the blend obtained by the TSC measurements remains almost constant for the different 
degradation times  (Fig. \ref{fig:Fig4}b). In this context it has to be mentioned that the determined trap density from the TSC spectra is only a lower limit of the actual trap 
density \cite{kadashchuk2005}. One reason is a possible recombination between electrons and holes after one of them got 
detrapped due to thermal activation. For degradation on synthetic air and simultaneous illumination we
\begin{figure}[htb]
	\centering
	\includegraphics[width=6.0cm]{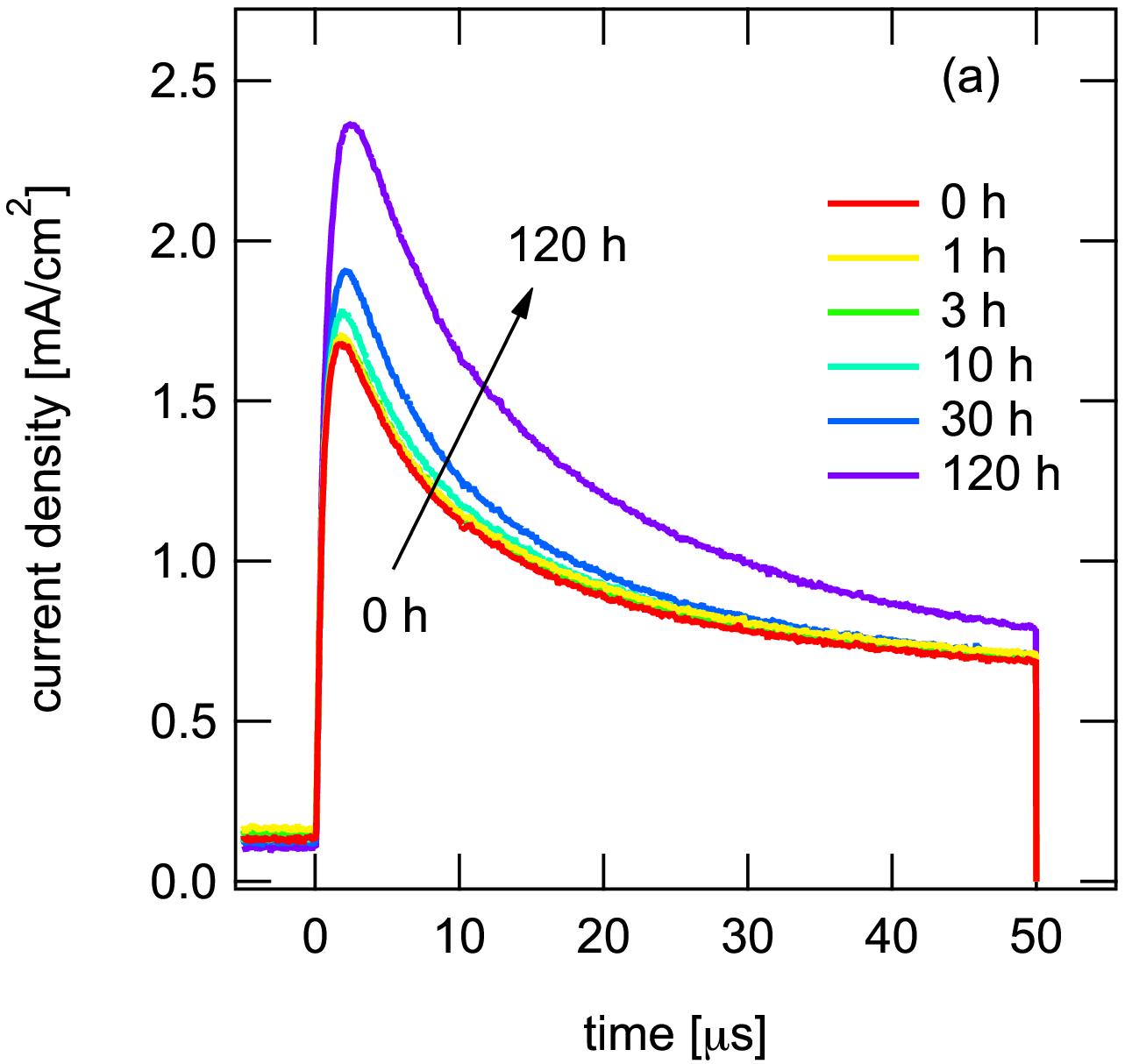}\\ 
	\includegraphics[width=6.3cm]{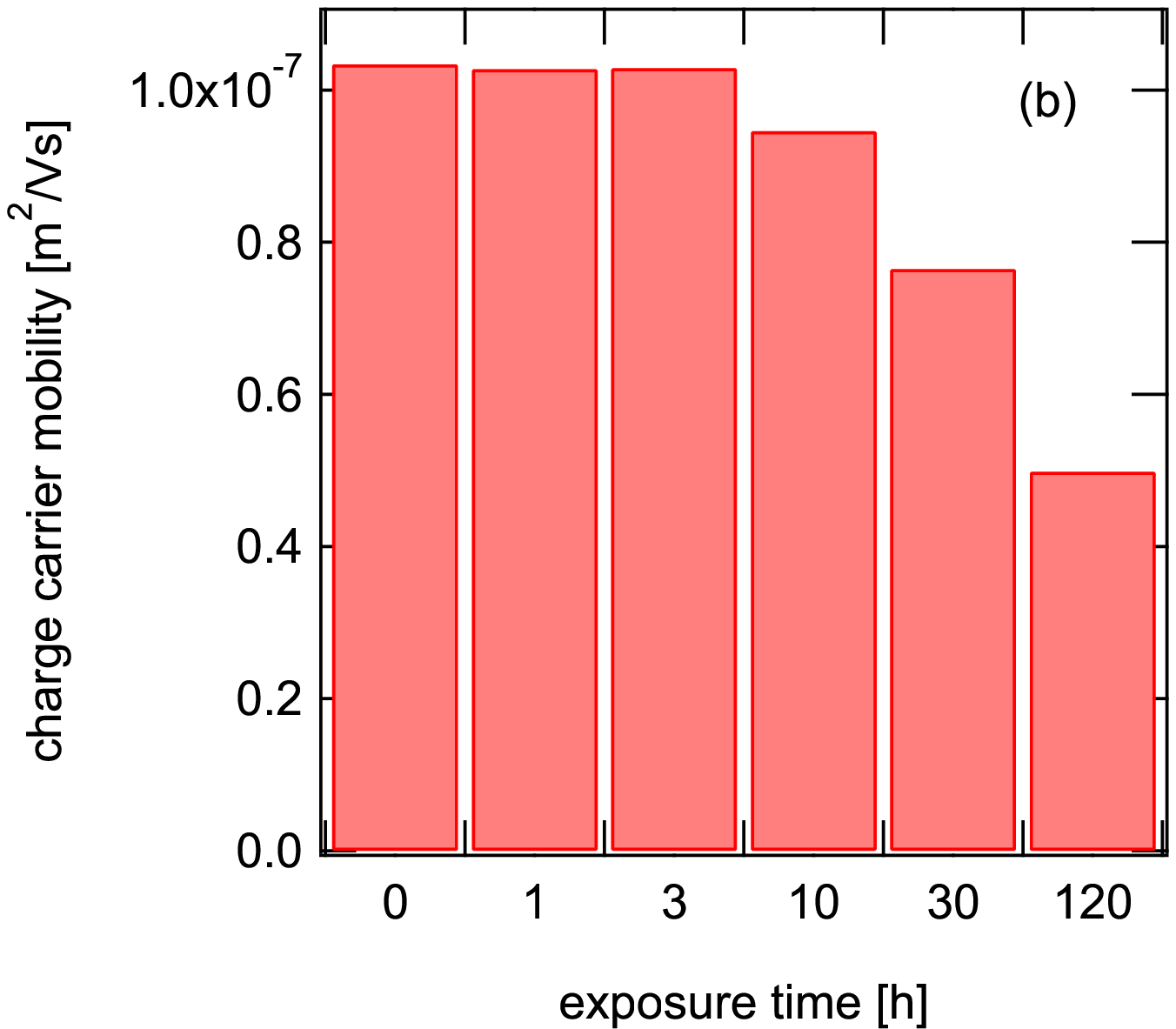}	
	\includegraphics[width=6.3cm]{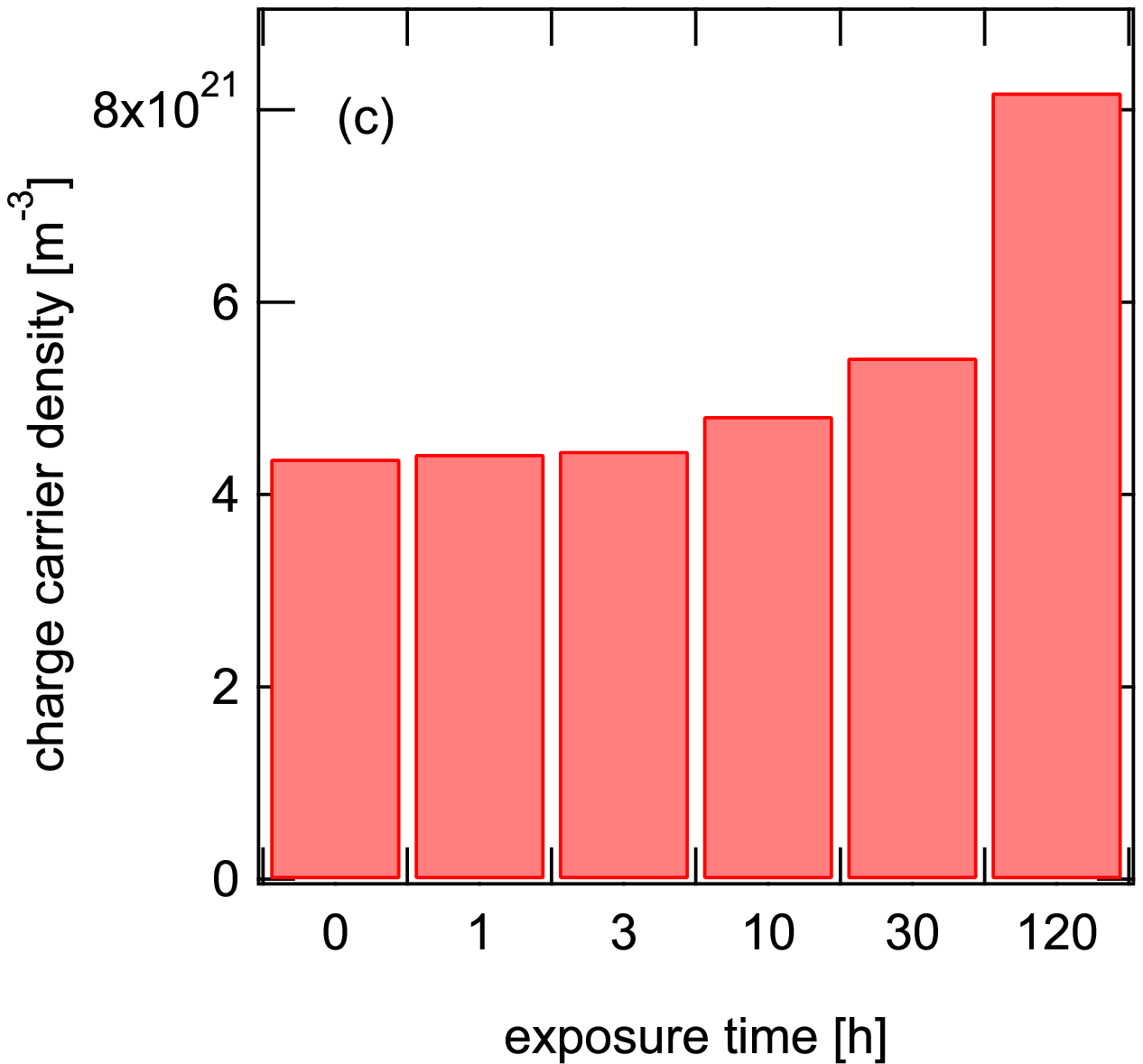}	
	\caption{CELIV measurements of P3HT:PCBM solar cells for different dark degradation times in synthetic air for offset bias of -0.3 V applied at the Ca/Al electrode (a) and the obtained mobilities (b) as well as the charge carrier densities (c). The charge carrier mobility slightly decreases with degradation time, 
whereas the charge carrier density shows an increase due to oxygen doping.}
	\label{fig:Fig5}
\end{figure}
 monitored almost the same behavior of TSC spectra, but on a faster timescale than during dark degradation.

To get further information about the influence of oxygen on the mobility and also the charge carrier concentration of the blend, we 
used the CELIV technique, which enables the simultaneous investigation of equilibrium charge carrier density and mobility by recording transient currents . 
The CELIV measurements for different degradation times in synthetic air in the dark are shown in Fig. \ref{fig:Fig5}a. The CELIV curves  exhibit a high asymmetry, especially for longer degradation times,  which might result from trapped charge carriers being extracted as the higher voltages at longer times promote emission from traps. This is in accordance with the increase of the deeper traps density as revealed by the TSC measurements. The mobility for different degradation times was calculated from the position of the current peak maximum according to reference \cite{deibel2009a}. Only a slight decrease of the 
mobility with degradation time can be observed (Fig. \ref{fig:Fig5}b), resulting in a decrease of about 50 \% after 120 
h in synthetic air.  
The equilibrium charge carrier density, extracted by the CELIV measurements, however, increases about a factor of two within a 
degradation time of 120 h in the dark (Fig. \ref{fig:Fig5}c). We assign these additional charge carriers to oxygen doping of P3HT, which was also reported for P3HT diodes  \cite{schafferhans2008} and P3HT field effect transistors \cite{meijer2003, liao2008}. We note that the concentration of charges measured by CELIV provides only a lower limit of the actual charge carrier density, since only charge carriers extracted within the time scale of the experiment (voltage pulse width typically 50 $\mu$s) are observed. Thus, the experimental number includes charges from comparably shallow traps which can be emitted from them and then be extracted; charges in deeper traps as well as the ones lost within the sample (by recombination) are disregarded. 
The degradation on synthetic air and simultaneous illumination shows similar results for the changes in mobility and charge carrier concentration, 
but again the degradation is accelerated to a timescale of minutes instead of hours.

\subsection{Simulation}
\label{Simulation}

In order to understand the influence of a decreased mobility as well as charge carrier doping on the solar cell parameters, we 
performed macroscopic simulations based on solving the differential equation system of the Poisson, the continuity and the drift--diffusion equations in one dimension \cite{deibel2008,deibel2009}. In the simulation we accounted a bimolecular non-geminate recombination rate according to Langevin \cite{langevin1903}. The charge carrier generation by light was assumed to create free polarons, neglecting any influence of electric field driven polaron pair dissociation. Furthermore the charge carrier mobilities of electrons and holes were set equal in the simulations. The used parameters are listed in Table \ref{tab:param}. 
 \begin{figure*}[htb]
	\centering
	\includegraphics[width=6.3cm]{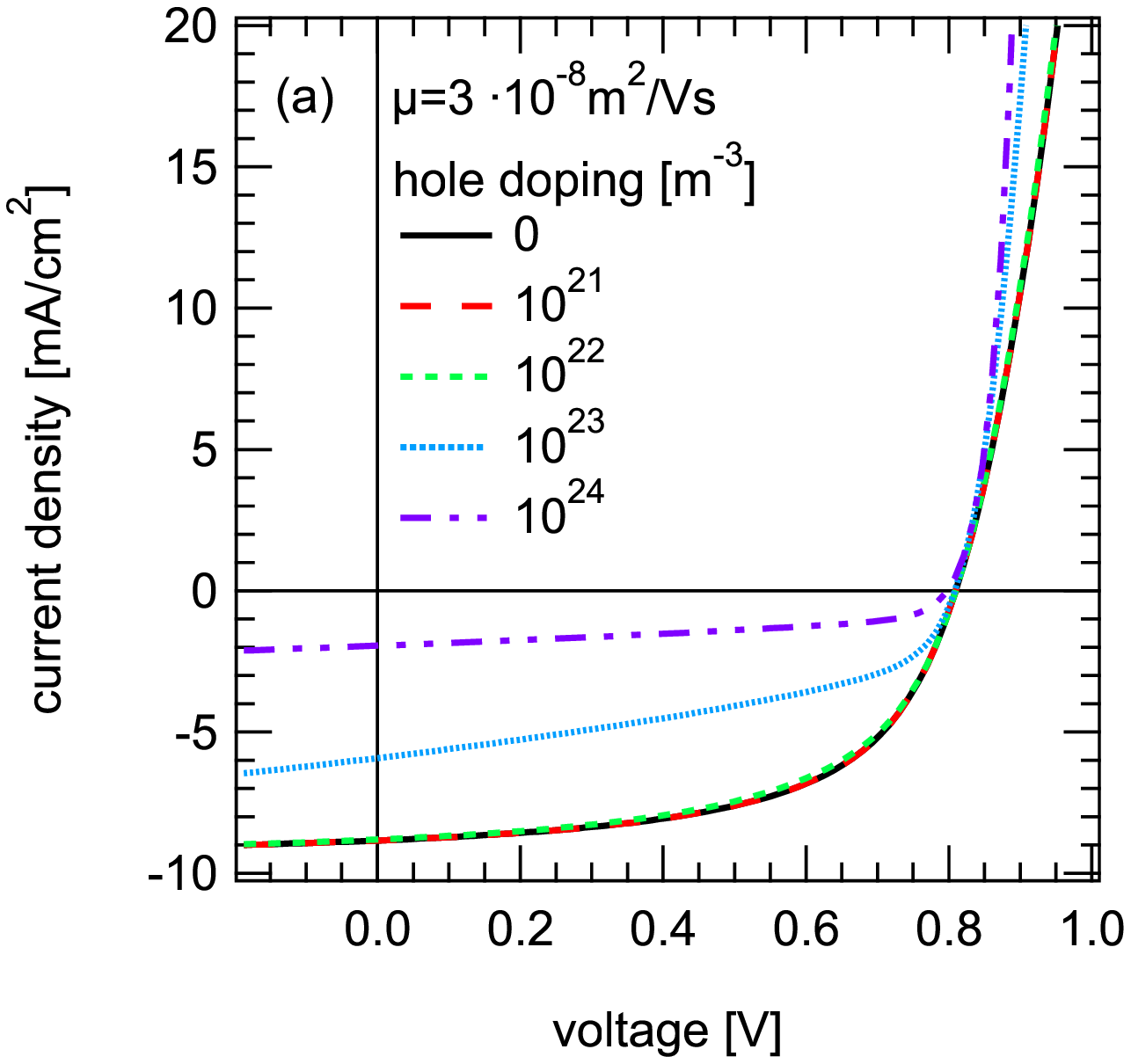}	
	\includegraphics[width=6.3cm]{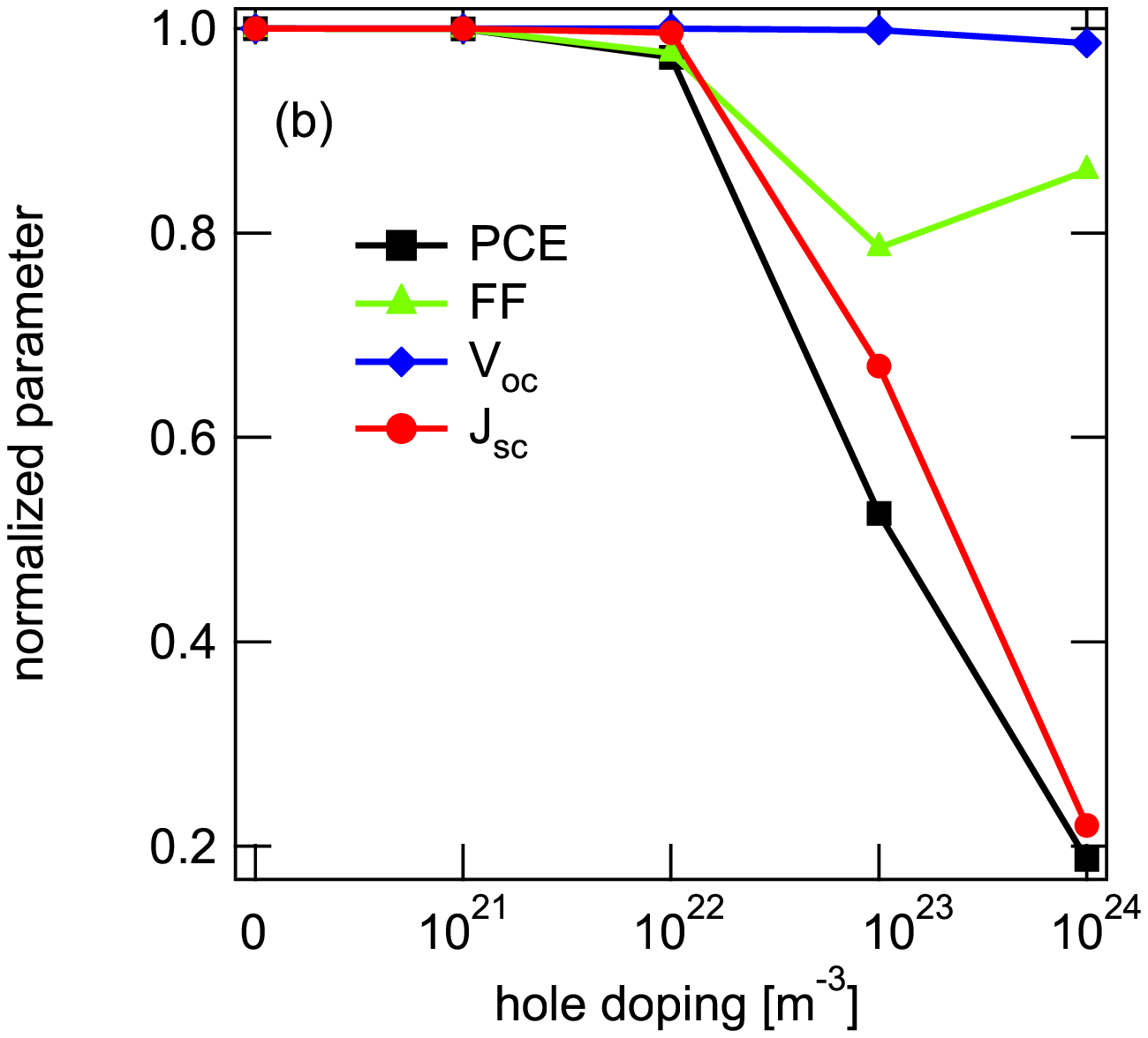}	
	\includegraphics[width=6.3cm]{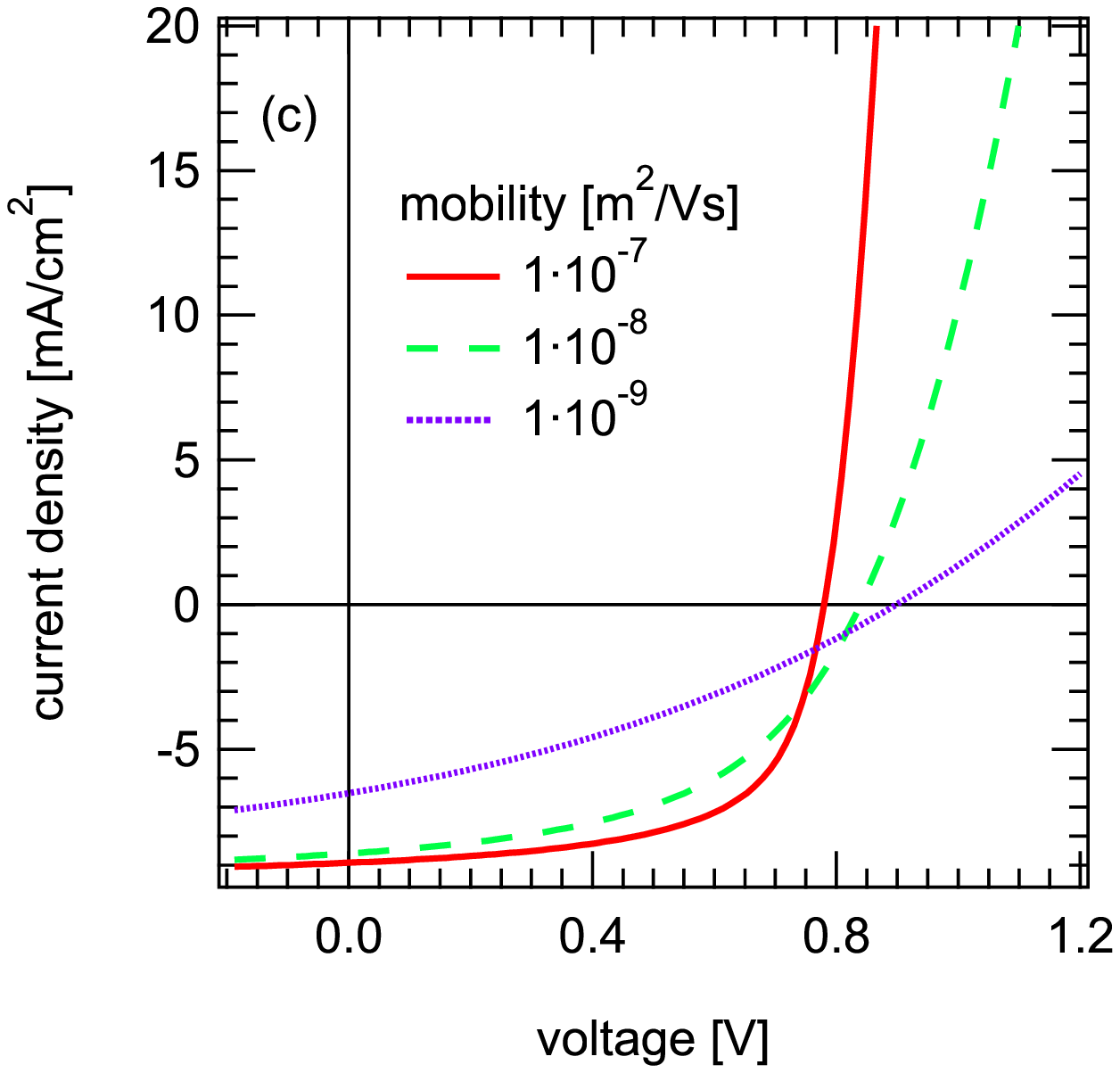}
	\includegraphics[width=6.3cm]{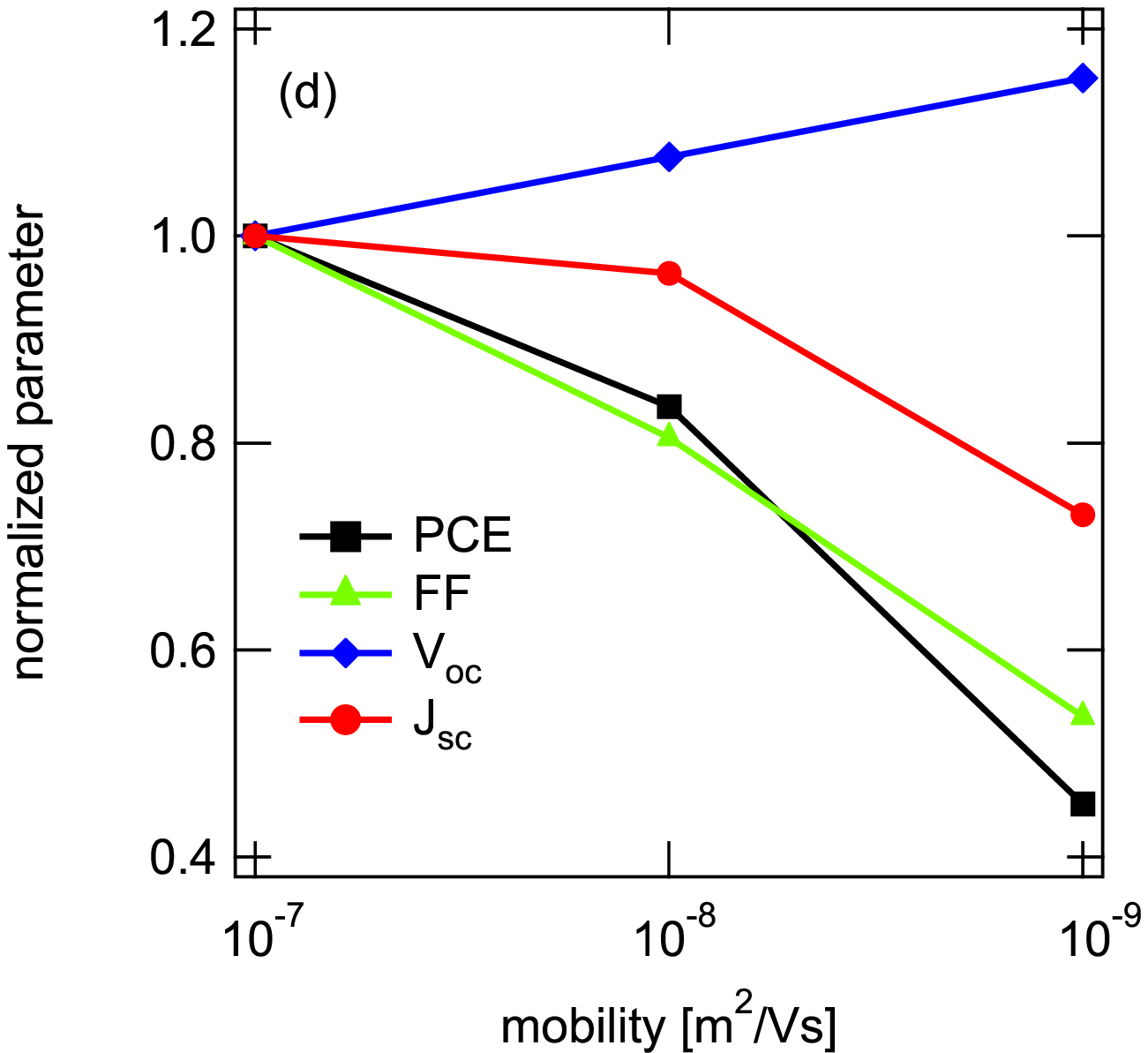}		
	\caption{Simulated IV-curves for different levels of hole doping (a) and the normalized values of the PCE, fill factor, open circuit voltage and short circuit current (b). An increased doping level results in a decrease of J$_{sc}$. A 
lowered charge carrier mobility (c) instead leads to a decrease of J$_{sc}$ and FF (d).} 
	\label{fig:Fig6}
\end{figure*}
Using this approach we investigated the effect of charge carrier doping by oxygen on the solar cell parameters. 
\begin{table}
	\centering
		\begin{tabular}{llll}
			\hline		
			parameter & symbol & value & unit \\

			\hline
			temperature & $T$ & 300 & K\\
			effective band gap & $E_g$ & 1.30 & eV\\
			relative dielectric constant & $\epsilon_r$ & $3.6$
& \\
			active layer thickness & $L$ & 100 & nm\\
			effective density of states & $N_c,~N_v$ & $
 8.0 \times 10^{25}$ & 1/m$^{3}$\\
			generation rate & $G$ & $6.0 \times 10^{27}$ &
1/m$^3$s\\
			injection barriers & $\Phi_a,~\Phi_c$ & $0.1$ & eV\\
			\hline
		\end{tabular}
	\caption{Parameters used for simulation.}
	\label{tab:param}
\end{table}

The simulated IV-curves for different hole doping levels are shown in Fig. \ref{fig:Fig6} a and the corresponding normalized solar cell parameters in  Fig. \ref{fig:Fig6} b. For the given parameter set, doping concentration up to 10$^{22}$ m$^{-3}$ does not change the PCE, whereas for higher 
concentrations the PCE strongly decreases due to a decrease of J$_{sc}$. FF and V$_{oc}$, however, remain almost unaffected. 
This behavior is almost independent of the mobility (within changes of one order of magnitude) used for the simulations. 
The decreased J$_{sc}$ due to doping can be explained by an increased bimolecular recombination probability. Since doping implies 
additional charges within the device, it leads to less band bending and therefore a reduced electrical field within the solar cell. As a 
consequence the extraction time for the charge carriers increases leading to a higher recombination probability and thus to a lower 
short circuit current.

The influence of the mobility on the solar cell performance is shown in Fig. \ref{fig:Fig6} c, d, revealing a loss of J$_{sc}$ and FF and a small increase of V$_{oc}$ for decreased mobilities. A lower mobility leads to a higher extraction time for the charge 
carriers and therefore to an enhanced bimolecular recombination probability, which results in the decreased FF. At the same 
time, the charge extraction becomes less efficient. The slightly increasing open circuit voltage is due to a lower Langevin recombination rate due to decreased mobilities. The higher charge carrier densities and their more balanced distribution reduce the internal electric field and therefore increase the open circuit voltage\cite{koster2005, cheyns2008}. If the field dependent polaron pair dissociation was also taken into account, it would mostly influence the FF and J$_{sc}$ resulting in an even stronger decrease with lower mobilities.

\subsection{Discussion}
\label{Discussion}

The influence of oxygen on P3HT:PCBM solar cells results in the case of dark degradation in a loss of J$_{sc}$, whereas 
photodegradation results in a decrease of all solar cell parameters. Further, the degradation due to oxygen is accelerated by light. A similar behavior was recently reported for inverted P3HT:PCBM solar cells \cite{seemann2009}, where the efficiency yields 
almost no change for dark degradation within the investigated timescale of 1h, whereas a strong loss for photodegradation within 
120 min is observed.
The loss in J$_{sc}$ can be explained by charge carrier doping, as we demonstrated by using macroscopic simulations. This 
interpretation is in accordance with the CELIV measurements, which show an increasing charge carrier concentration with 
degradation time. We attribute the increased charge carrier concentration to oxygen doping of P3HT \cite{schafferhans2008, meijer2003, liao2008}. The difference between the doping levels in the simulations and these determined by CELIV can be explained by the fact that the extracted charge carrier density by CELIV is only a lower limit of the actual charge carrier concentration, as mentioned before. The actual charge carrier density can be significantly higher. 
Furthermore, the accelerated increase of the charge carrier density in the case of photodegradation compared to the dark 
degradation is in good agreement with results of P3HT transistors, which also showed an expedited oxygen doping with light 
\cite{liao2008}.

In addition to an increased charge carrier concentration, CELIV measurements reveal a decrease of the charge carrier mobility for 
the P3HT:PCBM blends due to oxygen. A decreasing mobility with degradation time is also reported for pure P3HT 
\cite{schafferhans2008}, but in contrast to pure P3HT, where the mobility decreases about two orders of magnitude within 100 h 
\cite{schafferhans2008}, only a slight decrease in the blend---by about 50 \%---can be observed. This 
difference could be due to a stabilization effect of PCBM for P3HT in analogy to findings for MDMO-PPV (poly(2-methoxy-5-(3',7'-
dimethyloctyloxy)-1,4-phenylene-vinylene)):PCBM solar cells \cite{neugebauer2000}. The decreased mobility could also result from 
a degradation of PCBM due to oxygen. Decreased electron mobilities for C60 based field effect transistors after oxygen exposure have already 
been reported \cite{tapponnier2005, koenenkamp1999}. Since electron and hole mobilities can not be discriminated by CELIV, it 
is not possible to resolve which carrier mobility decreases.

A lower mobility leads to a decrease of FF and J$_{sc}$, as revealed by macroscopic simulations. Although our CELIV 
measurements show a slight decrease in mobility with degradation time, there are some facts indicating that the decreased mobility 
is not the origin of the FF loss in the case of photodegradation. First of all the experimentally observed mobility reduction is too limited to have a strong impact on the FF. Furthermore, the mobility decreases in the same range for both dark and illuminated degradation, although the degradation in the illuminated case occurs on a faster timescale. In contrast, the loss 
in FF is only observed for photodegradation. Additionally, we experimentally observed also a decrease of V$_{oc}$ for 
photodegradation, whereas a decrease in mobility should result in an increase of V$_{oc}$ as exhibited by the simulation (Fig. 
\ref{fig:Fig6} d). Therefore, the decreased mobility is not the reason for the FF loss.

Another important factor which can negatively influence the solar cell performance are electronic trap states, since they lower the 
mobility, disturb the internal field distribution and can act as recombination centers. 
Exposure of P3HT:PCBM solar cells to oxygen results in a rise of the density of deeper traps for both dark and illuminated 
degradation. On the other hand, the main peak of the TSC slightly decreases due to oxygen exposure, which is in contrast to the 
observed increase of the main TSC peak of pure P3HT \cite{schafferhans2008}. For comparison, the trap density in pure P3HT 
increases almost by a factor of three within 100 h under oxygen exposure \cite{schafferhans2008}.
In this context we want to mention again that the determined trap density from the TSC spectra is only a lower limit of the actual trap 
density \cite{kadashchuk2005}.

The appearance of deeper traps due to oxygen exposure could also be the reason for the observed loss in fill factor, since they can 
disturb the internal field distribution and act as recombination centers, as mentioned before. Further investigations concerning the 
influence of deep traps on the solar cell performance have to be done, like the implementation of deep traps within the 
macroscopic simulation. Other crucial factors concerning FF and V$_{oc}$ we want to address, are imbalanced electron--hole 
mobilities (resulting in the formation of space charges)  and a field dependent photogeneration. These two factors have not been 
accounted so far and will be addressed in the future to completely clarify the origin of the loss of FF and V$_{oc}$ in the case of 
oxygen induced degradation under simultaneous illumination.

\section{Conclusions}
\label{Conclusions}

We investigated the influence of oxygen on unprotected P3HT:PCBM solar cells by controlled exposure to synthetic air. Two different degradation conditions were used: in the dark and under simultaneous illumination. Exposure of the solar cells to synthetic air in the dark results in a loss of J$_{sc}$ of about 60 \% within 120 h. In contrast, simultaneous illumination during oxygen degradation results in a loss of all solar cell parameters, yielding an efficiency loss of about 30 \% within only 3 hours. Thus, the degradation of the P3HT:PCBM solar cells is strongly accelerated by light. 

CELIV measurements revealed an increased charge carrier concentration after oxygen exposure for dark, as well as 
photodegradation. We attributed these additional charge carriers to oxygen doping, which is also known for pure P3HT. With the aid 
of macroscopic simulations we have shown that doping of the solar cells is the origin of the loss in J$_{sc}$ for both degradation 
conditions. 

Another impact of oxygen exposure on P3HT:PCBM solar cells is a slight decrease of the charge carrier mobility, as also revealed 
by CELIV measurements. Although a decreased mobility may result in a loss of the fill factor, as shown by macroscopic simulations, the experimentally observed mobility decrease is too small to be the origin of the FF drop in the case of photodegradation.

In addition to an enhanced charge carrier concentration, oxygen induced degradation results in an increase of the density of 
deeper traps, as was revealed by TSC measurements. These deep traps could be the origin of the decreased FF and V$_{oc}$, although further investigations to correlate the trap density to the solar cell performance are required for a verification.

\section*{Acknowledgements}

The current work is supported by the Bundesministerium f\"ur Bildung und Forschung in the framework of the OPV Stability Project 
(Contract No. 03SF0334F). J.S. thanks the Elitenetzwerk Bayern for funding. A.B. thanks the Deutsche Bundesstiftung Umwelt for 
funding. C.D. gratefully acknowledges the support of the Bavarian Academy of Sciences and Humanities. V.D.'s work at the ZAE Bayern is financed by the Bavarian Ministry of Economic Affairs, Infrastructure, Transport and 
Technology. The authors thank Jens Lorrmann for fruitful discussion.





\bibliographystyle{model1-num-names}







\end{document}